# Tendencies, Dead-ends, and Promising Ways.
# From Interface Ideas to New Programs.

*Abstract.* The mechanism of communication between users and devices is called interface. From time to time changes in interface significantly improve our work with computers even without any serious changes in programs themselves. Main ideas in PCs interface were introduced many years ago and since then there are no significant changes, while new devices show promising ways by using direct manipulation of screen objects. Users' direct action with all the screen objects of our ordinary PCs turns standard screens into touch screens of very high resolution and not only changes the interface of familiar programs but creates the new type of programs: user-driven applications.

35 and more years ago the computer world consisted of two parts: hardware and software. The first one was very solid and only by its view caused an admiration from visitors of computer centers. The software was something unreal and not too many people knew how to deal with this non-material substance. Software, together with radio waves and subatomic particles, belonged to the set of invisible and untouchable things whose only evidence of existence was very helpful results of their work.

With the introduction of PCs, the number of computer users increased and started to grow at astonishing rate. For many users the hardware details are not interesting at all, but there are still two items in which every user is really interested: what computer can do for me (software) and how to communicate with the needed programs (interface). Interface is often looked at as some part of software design. Incorporation of interface into the main bulk of code and separation of interface from other parts are among the main issues of programming history. Tendencies in interface design affect much wider area than interface itself.

30 years ago Apple's Lisa demonstrated the graphical interface with such features which since then became standard for personal computers. They were quickly copied in Windows and under the patronage of this system conquered the world. Moving icons around the screen, moving and resizing windows, opening menus… Throughout the years there were many ideas in interface design but the basic things used on PCs are still the same.

At nearly the same time the mobile phone systems began to spread. Soon mobile phones also used the same idea of menus and submenus but small screen sizes and low resolution put a constraint on its use. Throughout the last years we see technical revolution in small devices and this caused a real revolution in their interfaces. Now technical characteristics of small devices bring them much closer to personal computers, but their designers are smart enough not to copy the interface from PC. Whenever possible, the touch screens are used or programs that simulate the touch screens. The system of available commands is not too wide and the main instrument – a finger – is not too accurate but good enough to organize the direct manipulation of all the screen elements. User can click an object to open the associated program, scroll the surface, change the view scale, or push an object in the direction of its needed move. It is definitely a small subset of what can be done on PCs, but there is one common feature in all available actions: users deal directly with the screen objects and this became the main interface idea on all portable devices. On all portable devices … except personal computers! Though the idea of direct dealing with the screen objects was introduced on PC nearly 30 years ago, it was implemented only partly, for a very limited set of objects, and was never improved since then.

There are two ways for a person to deal with objects of the surrounding world: it can be a direct action or some interpretation by another person or an object (an instrument). The direct action is often more effective. Whenever possible, people prefer direct actions. Smart developers understand this, but sometimes they don't see the way to implement the best solution. In other cases the decision is made not by clever developers but by managers and marketing department; as a result, huge resources are spent on constructing a perfect highway into the dead-end.

Users of Windows and similar multi windows operating systems deal with computer programs at two different levels. The upper one can be called the level of operating system because at this level the system itself provides everything. Since the beginning of multi windows systems and up till now, users see only two types of objects at this level: there are icons and there are rectangular areas associated with applications. Icons are small non-resizable rectangles which can be moved around the screen and placed anywhere. Rectangular areas associated with programs are usually much bigger than icons. These areas can be moved and resized in the same easy way: press – move (resize) – release. At this level mouse and its cursor are used to turn an ordinary screen into a touch screen: user deals directly with any object and does exactly what he (user!) wants to do. This works perfectly but unfortunately it works <u>only at the upper level</u>.

When an application starts, a rectangular area appears on the screen; this area can be filled with an arbitrary set of controls and graphical objects. Controls were mostly developed many years ago and since then only few new controls were added to the list. The set of graphical objects is unlimited as it is produced by imagination and skills of programmers from around the world. While working with a program, users have to deal with the objects shown inside the application's area; this is another (inner) level of communication. From the very beginning, controls had some features for direct communication with them: users can type a text, set a check mark, select line(s) in a list, or click a button. These are very useful features, but don't try to find among them any methods for direct moving or resizing of controls by users. Such methods do not exist.

Operating systems never proposed any methods for direct communication with arbitrary graphical objects. I think that developers of those systems didn't know how to solve this problem. It is not an easy task. Rectangular areas of the upper level have only straight horizontal and vertical borders. It is easy enough to detect the presence of the mouse cursor in the vicinity of such line, so it is easy to detect the grabbing of the line by mouse if the mouse button is pressed at such moment. Calculation of new position for horizontal and vertical straight lines according to the movement of the mouse is also not a problem. Now think a bit about the problems you have to solve when objects of arbitrary shape are involved in forward moving, resizing, and rotation. Even the straight borders can be placed at any angle, but you also have curved borders. In addition to solid objects of arbitrary shape, you have objects with holes of arbitrary shape; underlying objects must be detected through those holes.

There are several programs in which the moving of some graphical objects is a mandatory thing; *Paint* is maybe the most widely known of such programs. It is not an easy task to develop an algorithm for moving objects of some set in your own program which is distributed as an executable file. It is not easy, but for skillful programmers it is possible. *Paint* is one example; there are several others; I did the same thing in one of my programs many years ago.

It is much more complicated task if you need to distribute some library knowing that other developers are going to use the methods from this library to move and resize screen objects of arbitrary shape. There is always the question of usability: if the proposed methods are too complicated then they are not going to be used.

Thus, around 30 years ago the movability and resizability of objects was introduced at the upper level of multi windows systems. It was a very limited movability because only the objects of one primitive shape – rectangles – were involved, but from users' point of view this is a perfect algorithm because it is very easy in use and works for all the involved elements.

As I said, developers of operating systems proposed a whole set of controls for design of applications but never proposed any algorithm for moving the inner elements of these applications. Were there libraries from which developers could take such algorithms and methods? NO! Otherwise you would see the use of such methods long ago and in many applications. It is strange, but many teachers on user interface think that such algorithm of turning any screen object into movable was demonstrated many years ago in the Morphic system. I can't find the source of this strange rumour, but it is definitely a mistake. There is a well known paper about Morphic system written by its authors [1]. There is not a single word about the movability of objects but there is straight explanation that elements can be placed only on one of the predetermined positions in relation to other elements (page 22, second column, paragraph at the bottom). This is a classical case of adaptive interface and nothing else; there are several choices prepared and coded by developers and available to users. Authors honestly wrote that even this was not a new feature but such positioning was used in earlier systems. It is a mystery why some "specialists" continue to insist that there was movability of objects in Morhic. Don't they see the difference between declaring a position on one of predetermined places and free moving of any object around the screen? These are two different worlds.

What are the main ideas of interface design on PC? For the last 30 years there is only one religion – *adaptive interface*. It is impossible to develop a program with an interface which every user will like, especially, when there are thousands or millions of users. People are different and there can be users even with opposite demands. In order to fulfill different requirements, the adaptive interface was introduced. The main idea of adaptive interface was really brilliant: provide users with an easy instrument to adapt a program to their tasks. 20 – 25 years ago there were hundreds or thousands of papers and books on some features of adaptive interface; the best ideas are still used in a lot of programs. Adaptive interface became an axiom, a dogma, but deep inside it contains a fundamental flaw which is never publicly discussed: adaptive interface is based on the postulate that designers know better than any user what is really good for each and all situations. The adaptive interface is usually called friendly because it gives users a choice; however, any selection is made only among the possibilities which are considered as appropriate by developers; only such variants are allowed. The interface is not fixed any more; it gives users some choices but it is still fully controlled by developer.

When users are not satisfied with a set of allowed solutions, then a new set of possibilities is coded or another instrument to select among them is developed. The number of commands and possibilities increases; the interface becomes overwhelmingly sophisticated. This situation is perfectly described in the preface to [2]: "*You have to figure out how to cast what you want to do into the capabilities that the software provides. You have to translate what you want to do into a sequence of steps that the software already knows how to perform, if indeed that is at all possible. Then, you have to perform these steps, one by one.*" As a result, users often are not doing what they want to do but try to find how to ask an application to do something close to their need.

Developers try to design their programs in the best possible way and think a lot about the best presentation of the needed information on the screen. At the same time people use screens with different resolutions and choose the preferable size of the fonts; combinations of these parameters can make a mess of any perfectly thought out screen view and ruin any program. If users could easily move and resize the screen objects, such mess on the screen would never occur, but instead of producing such instrument for moving / resizing, developers introduced the *dynamic layout*; it began to spread somewhere 10 years ago.

The idea of dynamic layout is simple. Developer provides a good view, as he understands it; user can change the outer sizes of the form (dialog) and the program adapts all inner elements and the view to this changed size. Adaptation is done according to some algorithm developed beforehand and absolutely independent from users, so users have no control at all over the inner view. Adaptive interface gives some imitation of users' control over the view of a program; dynamic layout dismisses even this imitation. Instead of direct users' ruling over the screen elements, there is only developers' interpretation of users' actions. The adherents of dynamic layout declare its simplicity for users but prefer not to discuss all its negative sides. Users are cut from making any decisions about the view of the programs they have to work with. For me it looks like a classical computing slavery. Not long ago I got such words in a private letter from one of the MIT professors: "*Without dynamic layout, the end user would have to manually, one by one, resize and reposition the elements inside. So dynamic layout does confer usability benefits by making the user interface more efficient: one resize action by the user results in many automatic resizes and repositions of dependent objects.*" In the same way somebody can declare that because slaves are provided with food and shelter and don't need to think about these things, then slavery is the best form of social organization.

You like the idea of dynamic layout but you don't like to be compared with the slavery proponents? Then look at this problem from such point. What is the main goal of any program: to minimize users' efforts and to do whatever this program (or its developer) prefers or to do what user wants? Are you still going to insist that there are situations when dynamic layout is the best choice? Then forget about the theoretical user. YOU are the user of the program which does something of its own and not what you want. Do you still think that this program works correctly?

That is why I declare all the time: adaptive interface and especially dynamic layout is the perfectly constructed highway into the dead-end.

What do we see if we compare the tendencies in interface design on PC and on all other tiny, small, and not so small modern devices with any kind of screen? All the new devices which appear on the market every year and all new versions of the devices which appeared throughout the last several years work under the same general idea: "Only user decides what he wants to do and what he wants to see on the screen". Maybe the possibilities are not too wide but they increase with each new version. For all new versions the main rule of users' ruling is the same and users deal directly with all the screen objects.

Interface ideas and the main solutions on PCs are the same as they were demonstrated 30, 20, or at least 10 years ago. A lot of words are declared about the priority of users' requests but the motto is still the same: "You have to like whatever we give you". Decades ago there were hot discussions about the foolproof design of scientific and engineering programs. It was at the time when no other programs existed; many algorithms were coded and then distributed in the form of libraries to be used in different projects. For methods included into math libraries, the foolproof design is a mandatory thing; when the same ideas are applied to interface design, it is at least a mistake and in reality it is a pure nonsense.

Microsoft and other big companies are struggling for existence while users turn more and more from PCs to other devices which give them better and more comfortable interface. At the same time these big companies do not want to look into the core of problem. After its Golden Age, adaptive interface turned into dogma which prevents any further development. If you are a developer or researcher in computer science, you don't need to wait until Microsoft or somebody else will produce you some better instrument for program design. You can try what already exists and then decide for yourself if it is useful or not. If it is useful, then you have a choice of using the existing instrument or developing your own.

There are three components of new design: idea – instrument – results.

## Idea

The main idea is to give users the full control over WHAT, WHEN, and HOW to show on the screen. I call such programs *user-driven* because not developer but only user makes all the decisions.

When you develop a program of the standard type, you produce a toy. It can be a very sophisticated toy with a lot of possibilities and tuning features, but it is still a toy because no one can use it outside the set of features which you, as a developer, have already fixed somewhere inside the code. When you develop a user-driven application, you produce an instrument. It still has some main purpose (an analogue: nobody is going to wash dishes with a hammer), but you have to develop an instrument for some purpose and not a single restriction more. Users are free to use this instrument in any way they want.

Don't think about users of your programs as fools which must be prevented from doing a lot of things. These are the same people who use cellular phones and a lot of other devices. Somehow they manage those devices; maybe they can manage also your programs for PC? By the way, you are one of PC users. Do you want other developers to consider you a fool and thus develop the programs which you use as designed for fools?

## Instrument

There is one requirement for user-driven applications: all the screen objects must be movable. I am sure that there can be different algorithms to produce such results. Several years ago I have invented such algorithm. I published several articles about this algorithm and some results. Eventually I wrote a book [3] accompanied by a huge Demo program with a lot of different examples. All codes of this Demo program are available.

Programmers deal with a wide variety of screen objects and the algorithm has to work with arbitrary objects. There are basic features of this algorithm and there are many different cases which can be interesting for different developers. From time to time I publish small articles on some features of the main algorithm or about applying this algorithm to one or another group of objects [4 – 6]. There is nothing in these articles which is not discussed in the book, but each article covers some specific area.

I want to emphasize once more: my algorithm is not a mandatory thing for user-driven applications. It is only my instrument which I have invented and now use and demonstrate. You can use this instrument or any other. Regardless of the instrument you use to turn screen objects into movable / resizable, you develop user-driven applications with exactly the same rules that are formulated and discussed in my book. The results are independent of the algorithm.

Newton and Leibniz worked on the invention of calculus independently. Millions of people using calculus do not know this fact but they use calculus to achieve results in their areas of interest. You can use any algorithm and methods to turn objects in your programs into movable. Regardless of the algorithm, you'll get the same results.

## Results

For users, and we are all users of computer programs, the algorithms implemented inside the code are not important at all. Only the results matter. 40 – 50 years ago when the communication between people and big computers was so complicated that demanded professional training, only few people could do such job. PC significantly increased this number, but some features of work under DOS were still a deterrent for many users. Windows and similar systems took this barrier away. At the upper level of communication with PC we moved from requirement of special knowledge to the direct manipulation of the screen objects. That was a revolution in PC interface, but I'll call it mini-revolution because it was limited by this upper level. At this level only some preliminary steps of our work is done. The main work on PCs is done inside the innumerous applications. When at this inner level users will switch to the direct manipulation of all the screen objects, then the real revolution in our work with computers will happen. This process already started on cellular phones and other small devices. Now it's time for PC.

I am not going to include into this article even a single picture. All pictures are static. The main feature of user-driven applications is the movability of all the screen objects. A picture shows one fixed view of some program. Move from fixed interface or adaptive interface (it is simply a fixed set of views) to user-driven application is not a move from one view to another which developer considers as better one from one or another point. And it's not a revolution only because several possibilities are replaced by much wider (in reality, infinitive) set of possibilities. It is a revolution in our work with computers because the main role in such work goes from developer to user.

Any object in user-driven application can be relocated at any moment according to personal preferences of each particular user and only when this user wants some change. Nobody is going to downplay the importance of good design and every application will still require a well designed default view. Whenever user wants, he can return by a single click to this best result of designer's work, but user and only user makes a decision. The selection of default view is only one of numerous decisions which he can make. And whether to change anything in the view or not is also decided by user.

Words are only words. Before starting to add movability to objects in his own programs, any developer would like to see how it works in other programs. For this purpose, I included a lot of examples into the book and some of these examples are also demonstrated as stand alone programs.

How this movability can change the well known programs? There is an example of widely used *Calculator* which didn't change its view for more than 25 years. On one side, this is excellent when you don't need to waste your time trying to find familiar things. On the other side, are you sure that this familiar view is the best for you? In the version which I use from time to time, buttons for digits and operations are positioned absolutely differently from the classical view, but the program works as it has to do and calculates all the required expressions. What you, as a user, prefer: an application looking according to designer's taste or according to your preferences?

Many people are interested in finding information about their relatives and drawing a family tree. There are such programs. Maybe you are satisfied with the interface of one of them, but chances are high that you are not entirely satisfied but have to follow its rules. In my simple application *FamilyTree* I demonstrate that no rules of tree design have to be strict. The program helps but not dictates. With all elements movable and resizable, you organize the view which you want. With very high probability, for the same system of family connections another person will draw his family tree differently.

This is one of the things why the new idea in interface design – the movability of each and all screen elements – moves us from ideas in interface to new programs. Applications with adaptive interface are aimed to satisfy the hypothetic average user. User-driven applications have to satisfy every user. Companies don't need any more to spend millions of dollars on very aggressive advertising in an attempt to compel users to like the new version. Each user can change a new program in any way he wants and he will continue to change it occasionally depending on his mood, some changes of the task, weather outside, and many other things. The reason for each change does not matter at all; the main thing is that at each particular moment the view is changed in such a way as user wants. This rule is not for some abstract person. This is for you. You use a lot of programs and each of them will look exactly as you want. You don't like such world of applications?

Another feature of user-driven applications – advantages grow linearly or even faster with the growing complexity of applications. If you are a designer of an application with two – three – four screen elements (a lot of programs are of such simple type), you can position them in some way you like and force users to work with this variant. Anyway, users have nothing else to do if they need the results of this application. With two or three clicks at the design time, you can implement dynamic layout and then explain to your users that this is the best possible level of interface design and that they (users!) have to be happy with it.

Now consider a case of an application to see the view and analyse some math function(s). Pupils of the fifth or sixth grade start to do such analysis (forgive me if it happens earlier) and there are people who are doing such analysis for decades throughout their life. Such task has very wide variations: number of functions to view and compare, number of plotting areas, sizes and positions of those areas, and so on. No fixed design and no variant of adaptive interface can produce a good application for such task because there will be users not satisfied with any proposed view. In reality three things are needed from developer of such program: easy and obvious way to define a function, easy mechanism to select functions to be shown in one plotting area, and absolute users' control over the design of plotting areas, their positioning, and resizing. All these things are demonstrated in the *FunctionAnalyser* application.

Scientific and engineering applications are among the most complex programs. Math methods to be used in these programs are usually known at the beginning of design but there is a lot of uncertainty about the use of these methods and about the demonstration of results because these things often change throughout the research work. As a rule, users of each scientific or engineering program are much better specialists in particular area of research than the developers of the programs they use. As a result, in the programs with previously designed view, much better specialists have to do their work inside the limitations enforced on their work by significantly lesser specialists; this is the cause of never ending conflict between developers and users of such programs. Switch from the current programs to user-driven applications can either entirely end this conflict or move it to very low level. I have designed several programs of the new type for specialists in thermodynamics; some of these programs are demonstrated with the book. The reaction of researchers was immediate: from that moment they demanded only the programs of new type in which they controlled everything and made all the decisions. Researchers also asked to find some magic way for turning old programs into user-driven.

I mentioned several of examples; there are many of them in the application accompanying the book. All codes of this Demo application are available, so developers can estimate the easiness of this magic switch to movable screen elements. There is only one way to estimate the new things in programming – TRY them. As one very clever scientist formulated it years ago: "In the discovery of secret things, and in the investigation of hidden causes, stronger reasons are obtained from sure experiments and demonstrated arguments than from probable conjectures and the opinions of philosophical speculators". [7]